\documentclass[useAMS,usenatbib]{mn2e}

\usepackage{graphicx}

\begin{document}


\newcommand{\lya}{Ly$\alpha$}
\newcommand{\heii}{He\textsc{ii}}
\newcommand{\hii}{H \textsc{ii}}
\newcommand{\hei}{He\textsc{i}}
\newcommand{\ion}[2]{[\textsc{#1}]#2}
\newcommand{\ha}{\ensuremath{\mathrm{H}\alpha}}
\newcommand{\hb}{\ensuremath{\mathrm{H}\beta}}
\newcommand{\nii}{{\ion{N ii}{6564}}}
\newcommand{\sii}{{\ion{S ii}{6716,6731}}}
\newcommand{\siii}{{\ion{S iii}{9069,9532}}}
\newcommand{\oiii}[1][5007]{{\ion{O iii}{#1}}}
\newcommand{\oii}{{\ion{O ii}{3727}}}
\newcommand{\ewha}{\ensuremath{\mathrm{EW(\ha)}}}
\newcommand{\dewha}{\ensuremath{\Delta \mathrm{EW}(\ha)}}
\newcommand{\ssfr}{SFR/M$_*$}
\newcommand{\HII}{H\textsc{ii}}
\newcommand{\apjs}{ApJS}
\newcommand{\apj}{ApJ}
\newcommand{\aj}{AJ}
\newcommand{\araa}{ARA\&A}
\newcommand{\apjl}{ApJL}
\newcommand{\nat}{Nature}
\newcommand{\pasp}{PASP}
\newcommand{\aap}{A\&A}
\newcommand{\mnras}{MNRAS}

\def\ltsima{$\; \buildrel < \over \sim \;$}
\def\simlt{\lower.5ex\hbox{\ltsima}}
\def\gtsima{$\; \buildrel > \over \sim \;$}
\def\simgt{\lower.5ex\hbox{\gtsima}}


\title[C, N, O in the most metal-poor DLAs]{C, N, O Abundances in the 
Most Metal-Poor Damped Lyman alpha Systems\thanks{
Based in part on data obtained at the W.M. Keck Observatory,
which is operated as a scientific partnership among the California Institute 
of Technology, the University of California, and NASA, and was made possible
by the generous financial support of the W.M. Keck Foundation.
Based in part on UVES observations made with the European Southern 
Observatory VLT/Kueyen telescope at Paranal, Chile, obtained
in programme 078.A-0185(A) and from the public data archive.}
}

\author[M. Pettini et al.]
       {Max Pettini$^1$, Berkeley J. Zych$^1$, Charles C. Steidel$^2$, and 
        Fred H. Chaffee$^3$\\
         $^1$Institute of Astronomy, University of Cambridge, 
         Madingley Road, Cambridge CB3 0HA, UK\\
	 $^2$California Institute of Technology, Mail Stop 105-24, Pasadena, CA 91125, USA \\
         $^3$W. M. Keck Observatory, 65-1120 Mamalahoa Highway, 
         Kamuela, HI 96743, USA
      }

\date{Accepted ---; Received ---; in original form ---}
\pagerange{\pageref{firstpage}--\pageref{lastpage}}
\pubyear{2008}

\maketitle

\label{firstpage}

\begin{abstract}

This study focuses on some of the most metal-poor
damped \lya\ absorbers known 
in the spectra of high redshift QSOs,
using new and archival observations obtained 
with UV-sensitive echelle spectrographs
on the Keck and VLT telescopes. 
The weakness and simple velocity structure of 
the absorption lines in these systems allow us to
measure the abundances of several elements, and
in particular those of C, N, and O,  a group
that is difficult to study in DLAs
of more typical metallicities.
We find that when the oxygen abundance is less 
than $\sim 1/100$ of solar, the C/O ratio in high
redshift DLAs and sub-DLAs matches that of 
halo stars of similar metallicity and shows higher values 
than expected from galactic chemical evolution
models based on conventional stellar yields.
Furthermore, there are indications that at these
low metallicities the N/O ratio may also be above
simple expectations and may exhibit a minimum value,
as proposed by Centuri{\'o}n and her collaborators
in 2003.
Both results can be interpreted as evidence for 
enhanced production of C and N by massive stars
in the first few episodes of star formation, in our
Galaxy and in the distant proto-galaxies seen as 
QSO absorbers. The higher stellar yields implied
may have an origin in stellar rotation which promotes
mixing in the stars' interiors, as considered in some
recent model calculations.  We briefly discuss the
relevance of these results to current ideas on the origin
of metals in the intergalactic medium and 
the universality of the stellar initial mass function.

\end{abstract}


\section{Introduction}
\label{sec:introduction}

The neutral hydrogen clouds which give rise to 
damped Lyman alpha systems 
(DLAs; Wolfe, Gawiser, \& Prochaska 2005) 
in the spectra of high redshift QSOs provide important 
information on the early chemical evolution of galaxies,
complementing that obtained from 
more traditional studies of metal-poor
stars and H\,{\sc ii} regions in the Milky Way 
and nearby galaxies (Pettini 2004; Henry \& Prochaska 2007).
The relatively simple physics of interstellar line formation 
(Str{\"o}mgren 1948) makes the determination of 
ion column densities and corresponding element abundances 
straightforward, and the rich variety of resonant transitions 
at rest-frame ultraviolet wavelengths allows a detailed
picture of the chemical composition of many DLAs to 
be assembled (e.g. Prochaska, Howk, \& Wolfe 2003; Rix et al. 2007).
The elements carbon, nitrogen and oxygen have, 
however, been comparatively little studied so far in DLAs, 
despite the key role they play in stellar nucleosynthesis.
C and O are abundant elements with intrinsically strong
atomic transitions; in combination, these two factors 
result in resonance absorption lines which are strongly
saturated---and thus unusable for abundance 
determinations---even at the low metallicities of most DLAs.
The problem is particularly acute for C. For O some
weaker transitions are available at far-UV wavelengths;
these, however, can only be accessed from the ground 
at redshifts $z \simgt 2.5$. The absorption lines of N, on the
other hand, tend to be weak and are often blended with
intergalactic Ly$\alpha$ forest lines; nevertheless, the abundance
of N in DLAs has been the subject of several studies since
attention was first drawn to it by Pettini, Lipman \& Hunstead (1995). 

\begin{table*}
\centering
\begin{minipage}[c]{1.0\textwidth}
    \caption{\textsc{Journal of Observations}}
    \begin{tabular}{@{}lcrrccccrr}
    \hline
    \hline
   \multicolumn{1}{c}{QSO}
& \multicolumn{1}{c}{$V$ or $g^{\rm a}$} 
& \multicolumn{1}{c}{$z_{\rm em}$}
& \multicolumn{1}{c}{$z_{\rm abs}$}
& \multicolumn{1}{c}{Telescope/}
& \multicolumn{1}{c}{Wavelength}
& \multicolumn{1}{c}{Resolution}
& \multicolumn{1}{c}{Integration Time}
& \multicolumn{1}{c}{S/N$^{\rm b}$}
& \multicolumn{1}{c}{$N$(H~{\sc i})}\\
   \multicolumn{1}{c}{ }
& \multicolumn{1}{c}{(mag)}
& \multicolumn{1}{c}{ }
& \multicolumn{1}{c}{ }
& \multicolumn{1}{c}{Instrument }
& \multicolumn{1}{c}{ Range (\AA)}
& \multicolumn{1}{c}{(km~s$^{-1}$)}
& \multicolumn{1}{c}{(s)}
& \multicolumn{1}{c}{ }
& \multicolumn{1}{c}{(cm$^{-2}$)}\\
    \hline
Q0913$+$072       & 17.1 & 2.785  & 2.61843  & VLT/UVES               & 3310--9280$^{\rm c}$     & 6.7   & 77\,500  & 60~~ & $2.2 \times 10^{20}$  \\
SDSS\,J1016+4040  & 19.5 & 2.991  & 2.81633  & Keck\,{\sc i}/HIRESb                  & 3190--5970$^{\rm c}$     & 7.5   & 15\,000  & 12~~ & $0.8  \times 10^{20}$  \\
SDSS\,J1558+4053  & 18.7 & 2.635  & 2.55332  & Keck\,{\sc i}/HIRESb                  & 3190--5970$^{\rm c}$     & 7.5   & 25\,200  & 18~~ & $2.0 \times 10^{20}$  \\
Q2206$-$199       & 17.3 & 2.559  & 2.07624  & VLT/UVES \&  & 3100--10\,000$^{\rm c}$  & 6.5   & 68\,500  & 100~~ & $2.7 \times 10^{20}$  \\
& & & &  Keck\,{\sc i}/HIRESr  & &  & & &   \\
    \hline
    \end{tabular}
    \smallskip

 $^{\rm a}$Magnitudes are $g$ for SDSS objects and $V$ for other QSOs.\\
 $^{\rm b}$Indicative signal-to-noise ratio at 5000\,\AA.\\
 $^{\rm c}$With some wavelength gaps\\
    \label{tab:obs}
\end{minipage}
\end{table*}

The relative abundances of C, N, and O at low metallicities
hold clues to the nature of the stars responsible for their 
production in the earliest stages of nucleosynthesis in galaxies,
and their measurement in DLAs has the potential of clarifying 
some unresolved questions raised by recent stellar work,
as discussed in detail later. The marked increase in the number
of known DLAs afforded by the Sloan Digital Sky Survey
(Prochaska, Herbert-Fort, \& Wolfe 2005) makes it possible
to search for damped systems with the characteristics which facilitate
the measurement of the abundances of C, N, and O, primarily
low metallicity and simple velocity structure. 
In this paper
we report new high resolution spectroscopic
observations of four such absorption 
systems,\footnote{Three DLAs and one sub-DLA, according
to the conventional distinction between the two at
neutral hydrogen column densities 
$N$(H\,{\sc i})\,$= 2 \times 10^{20}$\,cm$^{-2}$.} 
all with metallicities [O/H]\,$< -2$.\footnote{As usual,
[O/H]\,$\equiv \log($O/H)$_{\rm DLA} -  \log($O/H)$_{\odot}$.}
When combined with published measurements in two similar 
systems, this data set---although modest---sheds 
new light on the nucleosynthesis of C, N, and O  
in the low metallicity regime.

\begin{figure*}
  \vspace*{-1cm}
  \centering
  {\hspace*{-0.25cm}\includegraphics[angle=0,width=190mm]{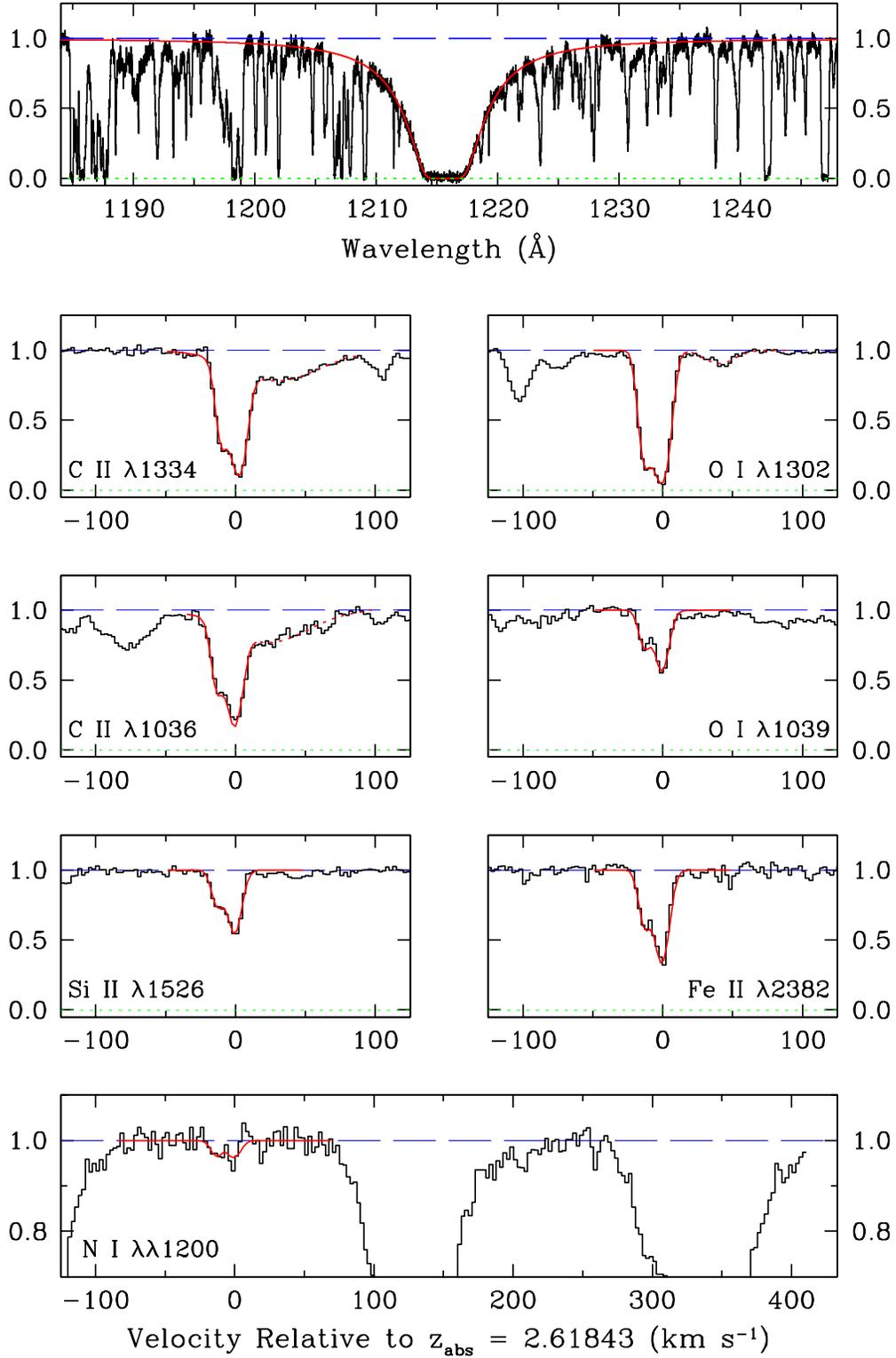}}
  \vspace{-3cm}
  \caption{
   Observed profiles (black histograms) and fitted Voigt profiles 
   (continuous red lines) of selected absorption lines in 
   the $z_{\rm abs} = 2.61843$ DLA in Q0913+072. 
   Dashed red lines indicate nearby absorption not attributed 
   to the DLA.  The $y$-axes of the plots show residual
   intensity; note the expanded $y$-scale of the bottom plot.
   The full list of absorption lines analysed in this DLA is
   given in Table~\ref{tab:Q0913_coldens}.
   }
   \label{fig:Q0913_montage}
\end{figure*}

\section{Observations and Data Reduction}
\label{sec:obs}
\subsection{Target Selection}
The four QSOs observed were selected based on the weakness
of metal lines in DLAs already known to be present in their spectra,
either from the trawl of the SDSS QSO sample by Prochaska et al. (2005)
or from earlier published studies. It has now become 
apparent that metallicity and strong metal line widths 
are correlated in DLAs (e.g. Ledoux et al. 2006a; 
Murphy et al. 2007a; Prochaska et al. 2007).
Thus, an efficient way to preselect for further study
the most metal-poor DLAs is to focus on DLAs in which the associated 
strongest metal lines, which are nearly always saturated, have
very low values of equivalent width---so low as to remain
undetected at the moderate resolution and signal-to-noise ratio
(S/N) of most SDSS spectra. In addition,
the simple kinematic structure of the absorbing gas
implied by the low equivalent widths simplifies
the abundance analysis compared with the more usual
occurrence of multiple absorption components spread
over 100--200\,km~s$^{-1}$.

Using this method, we have discovered
two of the most metal-poor DLAs, 
with oxygen abundance [O/H]\,$\simeq -2.4$,
in the spectra of the QSOs SDSS~J1016+4040 and
SDSS~J1558+4053; high resolution observations of these
systems are reported here for the first time.
We have also reobserved two QSOs with known
low-metallicity, narrow-line DLAs: 
Q0913+072 and Q2206$-$199; the new data 
add significantly to those already available
in the literature thanks to the very high S/N ratio achieved.

\subsection{Observations}

The observations were performed with the 
HIRES (Vogt et al. 1994) and UVES (Dekker et al. 2000)
echelle spectrographs on, respectively,
the Keck\,{\sc i} and VLT2 telescopes.
Relevant details are collected in Table~\ref{tab:obs}.
For the SDSS QSOs we used the refurbished HIRES 
with its UV-sensitive cross-disperser,
covering the wavelength range 3190--5970\,\AA.
The UVES observations of Q0913+072 and Q2206$-$199
used a variety of dichroics and wavelength settings
to cover essentially all of the optical wavelength
range from $\sim 3200-3300$\,\AA\ to $\sim 9000-10\,000$\,\AA.
The resolution is FWHM\,$\simeq 6 - 7$\,km~s$^{-1}$
for spectra recorded with either spectrograph.
The S/N ratio varies considerably along the spectrum,
particularly for the UVES data where a given portion
of the spectrum may have been covered with a number of
different instrument settings; in the penultimate
column of Table~\ref{tab:obs} we list an indicative
value of S/N at 5000\,\AA.

\subsection{Data reduction}

The reduction of the two dimensional images into
co-added one-dimensional spectra followed standard
procedures. The HIRES spectra were reduced with
the {\sc makee} pipeline software kindly maintained
by T.A. Barlow, while for the UVES spectra we used
the pipeline reduction provided 
by the European Southern Observatory but
with modifications which improve the accuracy
of the wavelength calibration and flux extraction
(see Murphy et al. 2007b).
These pipelines take
care of the initial stages of bias subtraction,
flat-fielding, wavelength calibration,
extraction of individual echelle orders,
and flux calibration.
The individual spectra of each QSO were averaged and the
echelle orders merged with purpose-built
software\footnote{Available from
http://astronomy.swin.edu.au/$\sim$mmurphy/UVES\_popler}
designed to maximise the signal-to-noise ratio
of the final spectrum which was binned
onto a linear, vacuum heliocentric wavelength
scale with approximately three bins per resolution
element.

\subsection{Abundance Measurements}\label{sec:abund_meas}

Absorption lines from each DLA were identified
and their profiles fitted with 
Voigt profiles generated by the VPFIT (version 8.02)
software package, as described by Pettini et al. (2002a)
and Rix et al. (2007).
Briefly, VPFIT uses $\chi^2$ minimisation to 
deduce the values of redshift $z$, column density $N$ (cm$^{-2}$), 
and Doppler parameter $b$ (km~s$^{-1}$)
that best reproduce the observed absorption line profiles.
VPFIT takes into account the 
instrumental broadening function
in its $\chi^2$ minimisation and error 
evaluation.\footnote{VPFIT is available from
http://www.ast.cam.ac.uk/\textasciitilde rfc/vpfit.html}
We used the compilation of laboratory
wavelengths and $f$-values by Morton (2003)
with recent updates by Jenkins \& Tripp (2006).

The DLAs observed are among those with the simplest
kinematic structure, generally consisting of only
two or three absorption components separated
by $\sim 10-20$\,km~s$^{-1}$; 
this fact, coupled with the high resolution and
generally high S/N of the spectra, made the
profile fitting procedure straightforward and
its results unambiguous. 
Another consequence of the unusually simple
line profiles is that the same set of `cloud'
parameters, $b$ and $z$, was found to reproduce 
well all the absorption lines considered in each DLA; 
naturally, lines arising from the same ground state of 
a given ion were fitted together to give
the best-fitting value of the column density $N$(X$^i$),
where $N$(X$^i$) is the ith ion stage of element X.
In each DLA, 
the column density of neutral hydrogen
was determined by fitting the
damping wings of the \lya\ line which are its most
sensitive measure; we also checked that
the resulting value of $N$(H\,{\sc i})
provides a good fit to the higher order Lyman lines
covered by our spectra.

Element abundances were deduced from the observed
ratios $N$(X\,{\sc n})/$N$(H\,{\sc i}), where
$N$(X\,{\sc n}) denotes the column density of 
the ion stage which contains the dominant fraction
of element X in H\,{\sc i} regions of the 
Milky Way interstellar medium (e.g. Morton et al. 1973).
For the elements considered here, these are
either the neutrals (O\,{\sc i}, N\,{\sc i})
or the first ions (C\, {\sc ii}, Al\,{\sc ii},
Si\,{\sc ii} and Fe\,{\sc ii}). However,
we have also measured the column densities
of higher ions stages, when their absorption
lines are present in our spectra, in order
to assess the corrections, if any, that need
to be applied
to our abundance determinations to account for
the possible contributions by ionised gas
to the absorption lines observed.
When element abundances are referred to the solar 
scale, we have used the compilation
by Asplund, Grevesse, \& Sauval (2005).

\section{Individual Absorption Systems}
\subsection{Q0913+072; DLA at $z_{\rm abs} = 2.61843$}

The metal-poor  DLA in front of this bright QSO has been known
since the survey by Pettini et al. (1997). It was subsequently studied
at higher spectral resolution by Ledoux et al. (1998) and Erni et al. (2006)
who considered it to be `the most metal-deficient DLA known'
and drew attention to its role as a probe of early nucleosynthesis.
As part of a programme to measure the primordial abundance
of deuterium, we re-observed this QSO with UVES in 2007.
Including a few spectra downloaded from the UVES data
archive, the total exposure time devoted to Q0913+072 is 77\,500\,s;
the corresponding signal-to-noise ratio near 5000\,\AA\ is S/N\,$\simeq 60$
(see Table 1).  

\begin{table}
\centering
    \caption{\textsc{Absorption Components of Low Ion Transitions in Q0913+072}}
    \begin{tabular}{@{}cccc}
    \hline
    \hline
  \multicolumn{1}{c}{Component}
& \multicolumn{1}{c}{$z_{\rm abs}$}
& \multicolumn{1}{c}{$b$}
& \multicolumn{1}{c}{Fraction$^{\rm a}$}\\
  \multicolumn{1}{c}{Number}
& \multicolumn{1}{c}{}
& \multicolumn{1}{c}{(km~s$^{-1}$)}
& \multicolumn{1}{c}{}\\
    \hline
1 & 2.61828    & 3.7  &  0.28 \\
2 & 2.61843    & 5.4  &  0.72 \\
    \hline
    \end{tabular}
    \smallskip

$^{\rm a}${Fraction of the total column density of Si\,{\sc ii}.}
    \label{tab:Q0913_cloudmodel}
\end{table}
\begin{table}
\centering
   \caption{\textsc{Ion Column Densities in Q0913+072, $z_{\rm abs} = 2.61843$ DLA}}
    \begin{tabular}{@{}llc}
    \hline
    \hline
  \multicolumn{1}{c}{Ion}
& \multicolumn{1}{c}{Transitions used} 
& \multicolumn{1}{c}{$\log N$(X)}\\
    \hline
H\,{\sc i}   &  1215                      & $20.34 \pm 0.04$ \\
C\,{\sc ii}  &  1036, 1334                & $13.98 \pm 0.05$ \\
N\,{\sc i}   &  1199                      & $12.29 \pm 0.12$ \\
O\,{\sc i}   &  1039, 1302                & $14.63 \pm 0.01$ \\
Al\,{\sc ii} &  1670                      & $11.78 \pm 0.03$ \\
Al\,{\sc iii} & 1854                      & $\leq 11.03$ \\
Si\,{\sc ii} &  1190, 1193, 1260, 1304, 1526 ~~   & $13.30 \pm 0.01$ \\
Fe\,{\sc ii} &  1608, 2344, 2374, 2382            & $12.99 \pm 0.01$ \\
     \hline
     \end{tabular}
     \smallskip
     \label{tab:Q0913_coldens}
\end{table}

A selection of the absorption lines in this DLA is 
shown in Figure~\ref{fig:Q0913_montage}.
The metal lines consist of two closely spaced
components; 
profile decomposition with VPFIT returns $b$-values of 3.7
and 5.4\,km~s$^{-1}$ and a velocity separation of 12.4\,km~s$^{-1}$
(see Table~\ref{tab:Q0913_cloudmodel}).
Table~\ref{tab:Q0913_coldens} lists the absorption lines
analysed and the corresponding values of column density;
the errors on $N$(X) are the formal values returned
by VPFIT.
Most of the metal lines in this DLA are weak;
in particular, from Figure~\ref{fig:Q0913_montage} 
it can be seen that both 
C\,{\sc ii}\,$\lambda 1334$ and C\,{\sc ii}\,$\lambda 1036$
are unsaturated (these lines are nearly always saturated in
DLAs), as is O\,{\sc i}\,$\lambda 1039$. 
Although both C\,{\sc ii} lines are partially
blended with broad absorption at longer wavelength
(modelled by VPFIT with two components), 
the impact of these contaminating features on the
determination of $N$(C\,{\sc ii}) is relatively minor:
a conservative estimate of the error on 
$\log N$(C\,{\sc ii}) deduced from several VPFIT trials 
is $\pm 0.05$. 

We have a definite ($4 \sigma$)
detection of N\,{\sc i}\,$\lambda 1199.55$
in both velocity components; the two other, weaker, members
of the N\,{\sc i}\,$\lambda \lambda 1199.55, 1200.22, 1200.71$
triplet are blended with \lya\ forest lines.

Turning to absorption lines from ionised gas,
C\,{\sc iii}\,$\lambda 977$, 
Si\,{\sc iii}\,$\lambda 1206$,
and Fe\,{\sc iii}\,$\lambda 1122$
are blended with \lya\ forest lines, while 
N\,{\sc ii}\,$\lambda 1084$ is undetected
(which is unsurprising given the weakness of the
N\,{\sc i} lines).
We do, however, have a tentative detection
of Al\,{\sc iii}\,$\lambda 1854$; given the 
weakness of this feature, we consider the column
density of Al\,{\sc iii} to be an upper limit.

\begin{figure*}
  \vspace*{-1cm}
  \centering
  {\hspace*{-0.25cm}\includegraphics[angle=0,width=190mm]{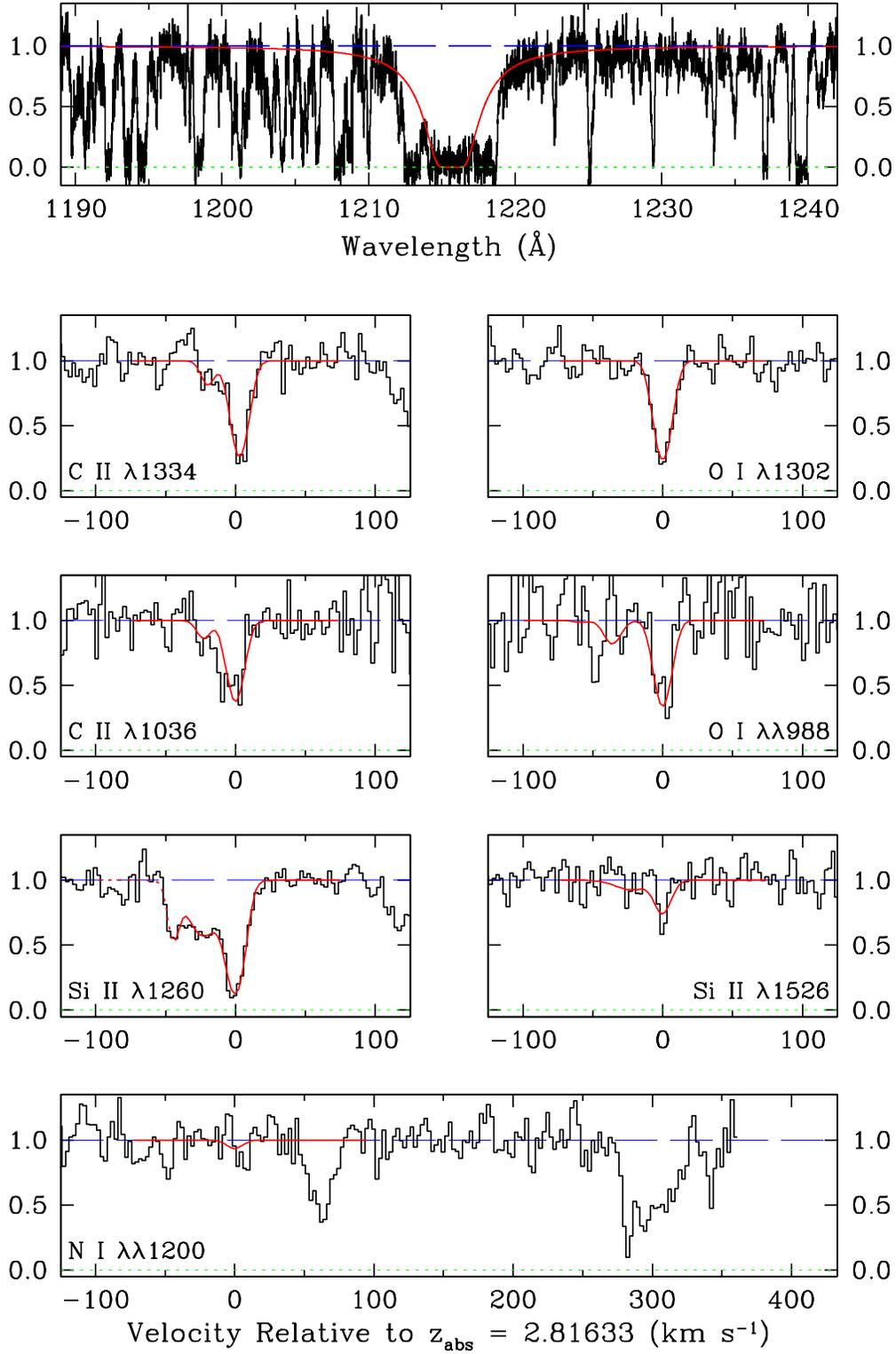}}
  \vspace{-3cm}
  \caption{
   Observed profiles (black histograms) and fitted Voigt profiles 
   (continuous red lines) of selected absorption lines in 
   the $z_{\rm abs} = 2.81633$ DLA in SDSS\,J1016+4040. 
   Dashed red lines indicate nearby absorption not attributed 
   to the DLA.  The $y$-axes of the plots show residual
   intensity. The full list of absorption lines analysed in this DLA is
   given in Table~\ref{tab:J1016_coldens}.
   }
   \label{fig:J1016_montage}
\end{figure*}

\subsection{SDSS\,J1016+4040; DLA at $z_{\rm abs} = 2.81633$}

This QSO is one order of magnitude fainter than Q0913+072.
We recorded its spectrum from 3190\,\AA\ to 5970\,\AA\
on the night of 2007 May 6 with the refurbished HIRES spectrograph
on the Keck\,{\sc i} telescope, using the UV cross-disperser
and UV sensitive CCD array. With a total exposure time of 15\,000\,s
we achieved a modest S/N\,$\simeq 12$ near 5000\,\AA.
As can be seen from Figure~\ref{fig:J1016_montage}, what appears
as a likely damped \lya\ line in the low resolution
SDSS spectrum breaks up into three distinct components
at the higher resolution of the HIRES data.
However, the central one 
of these three absorption lines does exhibit damping wings;
we deduce $N$(H\,{\sc i})\,$= (8 \pm 2) \times 10^{19}$\,cm$^{-2}$.
Such systems are sometimes referred to as sub-DLAs
since their column densities are lower than the historical definition
of a DLA as a system with $N$(H\,{\sc i})\,$> 2 \times 10^{20}$\,cm$^{-2}$
(Wolfe et al. 1986).

All the metal lines in this DLA are weak (see 
Figure~\ref{fig:J1016_montage}). The first ions exhibit 
two narrow velocity components with $b = 5.0$ and 6.9\,km~s$^{-1}$
separated by 23\,km~s$^{-1}$, with the latter 
accounting for $\sim 90$\% of the total column density
(see Table~\ref{tab:J1016_cloudmodel}).
The former, at $z_{\rm abs} = 2.81604$,
probably arises in ionised gas since it is not
detected in the O\,{\sc i}\,$\lambda 1302$
and $\lambda 988.77$ lines (the latter being the
strongest member of the O\,{\sc i}\,$\lambda \lambda 988.77, 988.65, 988.58$
triplet shown in Figure~\ref{fig:J1016_montage}).
Therefore, we have not included this component in the
values of $N$(C\,{\sc ii}) and $N$(Si\,{\sc ii}) attributed 
to the DLA in Table~\ref{tab:J1016_coldens}.
N\,{\sc i} is below our detection limit; similarly,
absorption lines from higher ion stages than those dominant
in H\,{\sc i} regions are either blended (e.g. Si\,{\sc iii}\,$\lambda 1206$),
or undetected. 

\begin{table}
\centering
    \caption{\textsc{Absorption Components of Low Ion Transitions in SDSS\,J1016+4040}}
    \begin{tabular}{@{}cccc}
    \hline
    \hline
  \multicolumn{1}{c}{Component}
& \multicolumn{1}{c}{$z_{\rm abs}$}
& \multicolumn{1}{c}{$b$}
& \multicolumn{1}{c}{Fraction$^{\rm a}$}\\
  \multicolumn{1}{c}{Number}
& \multicolumn{1}{c}{}
& \multicolumn{1}{c}{(km~s$^{-1}$)}
& \multicolumn{1}{c}{}\\
    \hline
1 & 2.81604    & 5.0  &  0.11 \\
2 & 2.81633    & 6.9  &  0.89 \\
    \hline
    \end{tabular}
    \smallskip

$^{\rm a}${Fraction of the total column density of C\,{\sc ii}.}
    \label{tab:J1016_cloudmodel}
\end{table}
\begin{table}
\centering
   \caption{\textsc{Ion Column Densities in SDDS\,J1016+4040, $z_{\rm abs} = 2.81633$ DLA}}
    \begin{tabular}{@{}llc}
    \hline
    \hline
  \multicolumn{1}{c}{Ion}
& \multicolumn{1}{c}{Transitions used} 
& \multicolumn{1}{c}{$\log N$(X)}\\
    \hline
H\,{\sc i}   &  1215                                      & $19.90 \pm 0.11$ \\
C\,{\sc ii}  &  1334, 1036                                & $13.66 \pm 0.04$ \\
N\,{\sc i}   &  1199.5                                    & $\leq 12.76    $ \\
O\,{\sc i}   &  1302, 988.77                              & $14.13 \pm 0.03$ \\
Si\,{\sc ii} &  1260,  1526 ~~~~~~~~                      & $12.90 \pm 0.05$ \\
    \hline
     \end{tabular}
     \smallskip
  \label{tab:J1016_coldens}
\end{table}

\subsection{SDSS\,J1558+4053; DLA at $z_{\rm abs} = 2.55332$}

\begin{figure*}
  \vspace*{-1cm}
  \centering
  {\hspace*{-0.25cm}\includegraphics[angle=0,width=190mm]{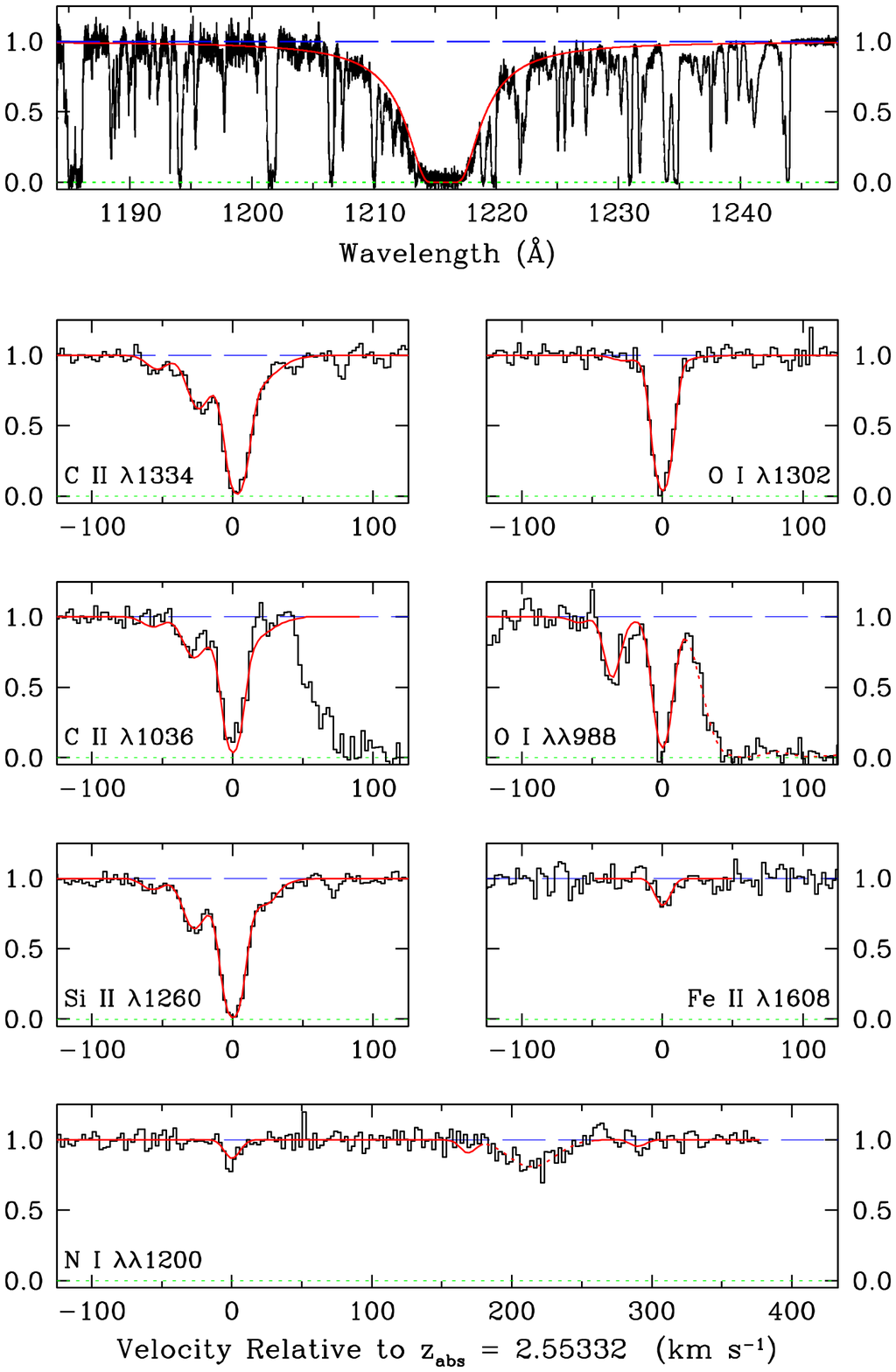}}
  \vspace{-3cm}
  \caption{
   Observed profiles (black histograms) and fitted Voigt profiles 
   (continuous red lines) of selected absorption lines in 
   the $z_{\rm abs} = 2.55332$ DLA in SDSS\,J1558+4053. 
   Dashed red lines indicate nearby absorption not attributed 
   to the DLA.  The $y$-axes of the plots show residual
   intensity. The full list of absorption lines analysed in this DLA is
   given in Table~\ref{tab:J1558_coldens}.
   }
   \label{fig:J1558_montage}
\end{figure*}

\begin{figure*}
  \vspace*{-1cm}
  \centering
  {\hspace*{-0.25cm}\includegraphics[angle=0,width=190mm]{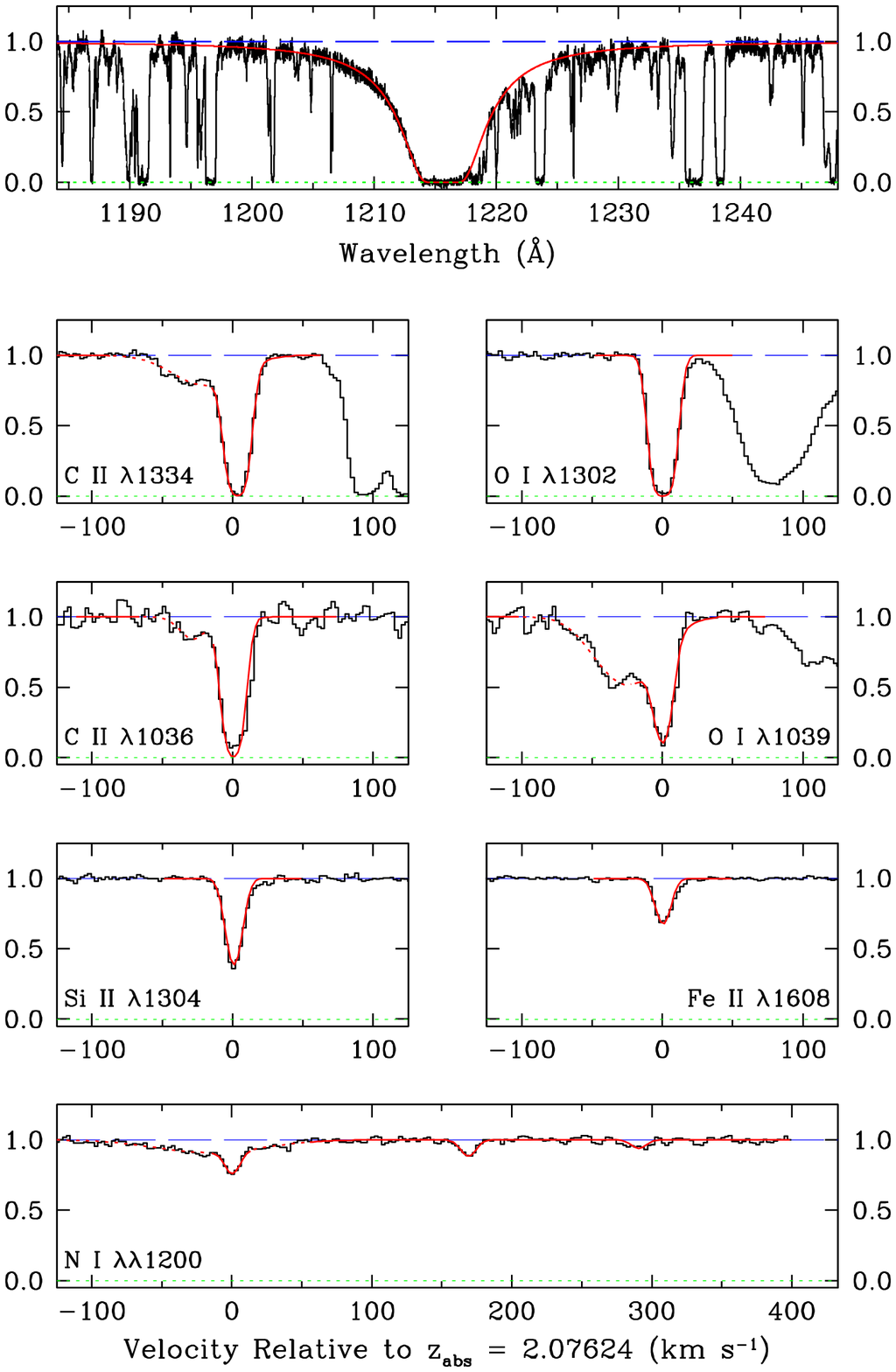}}
  \vspace{-3cm}
  \caption{
   Observed profiles (black histograms) and fitted Voigt profiles 
   (continuous red lines) of selected absorption lines in 
   the $z_{\rm abs} = 2.07624$ DLA in Q2206$-$199. 
   Dashed red lines indicate nearby absorption not attributed 
   to the DLA.  The $y$-axes of the plots show residual
   intensity. The full list of absorption lines analysed in this DLA is
   given in Table~\ref{tab:Q2206_coldens}.
   }
   \label{fig:Q2206_montage}
\end{figure*}

We observed this QSO with HIRES on the night of 2006 June 17,
achieving S/N\,$\simeq 18$ near 5000\,\AA\ with a 25\,200\,s
exposure. The echelle data confirm that the strong absorption
feature near 4320\,\AA\ in the discovery SDSS spectrum is
indeed a damped \lya\ line with 
$N$(H\,{\sc i})\,$= 2.0 \times 10^{20}$\,cm$^{-2}$
(see Figure~\ref{fig:J1558_montage}).

The neutral gas is concentrated in a 
single,  narrow ($b = 6.1$\,km~s$^{-1}$)
component at  $z_{\rm abs} = 2.55332$;
other, weaker, components are seen at
positive and negative velocities in C\,{\sc ii}
and Si\,{\sc ii} and probably originate in 
ionised gas (Figure~\ref{fig:J1558_montage}
and Table~\ref{tab:J1558_cloudmodel}).
The entries in Table~\ref{tab:J1558_coldens}
are for the neutral component only.
The absorption lines in this DLA are stronger
than in the previous two cases. Nevertheless,
we cover at least one unsaturated transition
for every species considered except possibly for
C\,{\sc ii}: both $\lambda 1334$ and $\lambda 1036$
are just saturated. Even for C\,{\sc ii}, however,
we can determine the column density with
confidence because: (i) the absorption profile (of the central, dominant
component) is well fitted by the same $b$-value which
fits best all the other lines; (ii) larger values of $b$ are
incompatible with the observed width of the line
which is resolved in our data. 
Note that while we cannot
exclude, in principle, that narrower components may be
present within the marginally saturated C\,{\sc ii}
line profiles,
their unrecognised presence would lead to an 
\emph{underestimate} of 
[C/H] and C/O, further strengthening the main
conclusion of our analysis of the behaviour of the 
C/O ratio at low metallicities, as discussed in Section~\ref{sec:C/O}.

As for the case of the DLA in SDSS\,J1016+4040, we are unable
to deduce reliable column densities of higher ionisation stages
than those listed in Table~\ref{tab:J1558_coldens} because the
relevant absorption lines are either blended, undetected, or
outside the wavelength range of our HIRES spectrum.

\begin{table}
\centering
    \caption{\textsc{Absorption Components of Low Ion Transitions in SDSS\,J1558+4053}}
    \begin{tabular}{@{}cccc}
    \hline
    \hline
  \multicolumn{1}{c}{Component}
& \multicolumn{1}{c}{$z_{\rm abs}$}
& \multicolumn{1}{c}{$b$}
& \multicolumn{1}{c}{Fraction$^{\rm a}$}\\
  \multicolumn{1}{c}{Number}
& \multicolumn{1}{c}{}
& \multicolumn{1}{c}{(km~s$^{-1}$)}
& \multicolumn{1}{c}{}\\
    \hline
1 & 2.55264    & 9.1  &  0.01 \\
2 & 2.55298    & 8.8  &  0.06 \\
3 & 2.55332    & 6.1  &  0.80 \\
4 & 2.55334    & 25.1 &  0.13 \\
    \hline
    \end{tabular}
    \smallskip

$^{\rm a}${Fraction of the total column density of Si\,{\sc ii}.}
    \label{tab:J1558_cloudmodel}
\end{table}
\begin{table}
\centering
   \caption{\textsc{Ion Column Densities in SDDS\,J1558+4053, $z_{\rm abs} = 2.55332$ DLA}}
    \begin{tabular}{@{}llc}
    \hline
    \hline
  \multicolumn{1}{c}{Ion}
& \multicolumn{1}{c}{Transitions used} 
& \multicolumn{1}{c}{$\log N$(X)}\\
    \hline
H\,{\sc i}   &  1215                                                   & $20.30 \pm 0.04$ \\
C\,{\sc ii}  &  1334, 1036                                          & $14.22 \pm 0.06$ \\
N\,{\sc i}   &  1199.5, 1200.2, 1200.7                        & $12.66 \pm 0.07$ \\
O\,{\sc i}   &  1302, 988.77, 988.65                           & $14.54 \pm 0.04$ \\
Al\,{\sc ii} &  1670                                                   & $11.92 \pm 0.06$ \\
Si\,{\sc ii} &  1190, 1193, 1260, 1304, 1526 ~~           & $13.32 \pm 0.02$ \\
Fe\,{\sc ii} &  1096, 1144, 1260, 1608,                       & $13.07 \pm 0.06$ \\
    \hline
     \end{tabular}
     \smallskip
  \label{tab:J1558_coldens}
\end{table}

\subsection{Q2206$-$199; DLA at $z_{\rm abs} = 2.07624$}

The metal-poor DLA in this QSO has been known since
the early echelle observations by Pettini et al. (1990)
and by Prochaska \& Wolfe (1997). Here we combine
spectra from the UVES and HIRES data archives 
to achieve S/N\,$\simeq 100$ at 5000\,\AA\ and
extensive spectral coverage from $\sim 3100$\,\AA\ to
$1\,\mu$m.

Even at this high resolution and signal-to-noise ratio
this DLA appears to consist of a single, narrow component:
the profiles of all the absorption lines from species which are
dominant in H\,{\sc i} regions are well reproduced 
by a model `cloud' with $b = 6.5$\,km~s$^{-1}$
(Figure~\ref{fig:Q2206_montage} and Tables~\ref{tab:Q2206_cloudmodel}
and  \ref{tab:Q2206_coldens}).
The availability of a number of transitions with differing
$f$-values for most ions ensures that the column densities
are well determined. The C\,{\sc ii} lines are just saturated but,
for the reasons outlined above, this is unlikely to affect
our conclusions concerning the relative abundances of C and O.
Thanks to its high S/N ratio and wide wavelength coverage, our
combined UVES+HIRES spectrum does cover a number
of transitions from ions that only occur in H\,{\sc ii} 
regions (N\,{\sc ii}, Al\,{\sc iii}, and Si\,{\sc iii};
see Table~\ref{tab:Q2206_coldens}); their column
densities are lower than those of the corresponding H\,{\sc i} region species.
Finally, it can be seen from Figure\,\ref{fig:Q2206_montage} that
we have a clear detection of all three lines of the
N\,{\sc i}\,$\lambda \lambda 1199.55, 1200.22, 1200.71$
triplet.

\begin{table}
\centering
    \caption{\textsc{Absorption Components of Low Ion Transitions in Q2206$-$199}}
    \begin{tabular}{@{}cccc}
    \hline
    \hline
  \multicolumn{1}{c}{Component}
& \multicolumn{1}{c}{$z_{\rm abs}$}
& \multicolumn{1}{c}{$b$}
& \multicolumn{1}{c}{Fraction$^{\rm a}$}\\
  \multicolumn{1}{c}{Number}
& \multicolumn{1}{c}{}
& \multicolumn{1}{c}{(km~s$^{-1}$)}
& \multicolumn{1}{c}{}\\
    \hline
1 & 2.07624    & 6.5  &  1.00 \\
    \hline
    \end{tabular}
    \smallskip

$^{\rm a}${Fraction of the total column density of Si\,{\sc ii}.}
    \label{tab:Q2206_cloudmodel}
\end{table}
\begin{table}
\centering
   \caption{\textsc{Ion Column Densities in Q2206$-$199, $z_{\rm abs} = 2.07624$ DLA}}
    \begin{tabular}{@{}llc}
    \hline
    \hline
  \multicolumn{1}{c}{Ion}
& \multicolumn{1}{c}{Transitions used} 
& \multicolumn{1}{c}{$\log N$(X)}\\
    \hline
H\,{\sc i}   &  1215                                  & $20.43 \pm 0.04$ \\
C\,{\sc ii}  &  1334, 1036                            & $14.41 \pm 0.03$ \\
N\,{\sc i}   &  1199.5, 1200.2, 1200.7                & $12.79 \pm 0.05$ \\
N\,{\sc ii}  &  1083                                  & $12.46 \pm 0.10$  \\
O\,{\sc i}   &  1302, 1039                            & $15.05 \pm 0.03$ \\
Al\,{\sc ii} &  1670                                  & $12.18 \pm 0.01$ \\
Al\,{\sc iii}&  1854, 1862                            & $11.59 \pm 0.04$ \\
Si\,{\sc ii} &  1193, 1260, 1304, 1808 ~~             & $13.65 \pm 0.01$ \\
Si\,{\sc iii}&  1206                                  & $\leq 12.86 $\\
Fe\,{\sc ii} &  1121, 1260, 1608, 2344, 2374, 2382, 2600    & $13.33 \pm 0.01$ \\
    \hline
     \end{tabular}
     \smallskip
  \label{tab:Q2206_coldens}
\end{table}

\section{Element Abundances}

Our abundance measurements in the four DLAs are collected
in Table~\ref{tab:abundances}. We have searched the literature
for other examples of DLAs (or sub-DLAs) where the 
C\,{\sc ii} and O\,{\sc i} absorption lines are unsaturated, so that 
the corresponding column densities can be determined, and
found only two other cases, also included in
Table~\ref{tab:abundances}, from the work by
Dessauges-Zavadsky et al. (2003) and P{\'e}roux et al. (2007).
Both of these cases are of somewhat lower hydrogen column density
and at higher redshifts than the four reported here but,
given the paucity of C/O measurements up to now, we shall
consider them in the present analysis.

As can be seen from Table~\ref{tab:abundances},
all four DLAs in the present work have metallicities
of less than 1/100 of solar.  
The DLA in front of Q0913+072 remains the most metal-poor
in [Fe/H], but its oxygen abundance of $\sim 1/250$ solar
is similar to those of the two newly discovered DLAs from the SDSS.
With [O/H]\,$\simeq -2.4$ and [C/H]\,$\sim -2.5$ to $-2.7$,
these DLAs have attained a level of metal enrichment
comparable to that typical of intergalactic \lya\ forest
clouds at the same redshifts (e.g. Simcoe et al. 2004). 
Although the source
of these intergalactic metals is still somewhat controversial
(see Ryan-Weber et al. 2006 and references therein),
it seems likely that such metal-poor DLAs have
experienced very little, if any, star formation \textit{in situ}. 

It is also interesting to note that no
DLA has yet been found with metallicity
less than 1/1000 solar, despite the fact that such metal-poor
stars do exist in the halo of our Galaxy.
In drawing such comparisons, however, it must be borne
in mind that such extremely metal-poor stars constitute 
a tiny fraction of the stellar populations of the Milky Way.
It may simply be the case that with only a few hundred
metallicity measurements in DLAs so far, we simply do
not have the statistics necessary to pick out such outliers
in the metallicity distribution.

Finally, we draw attention to the fact that in the four
systems in which this ratio could be measured,
[O/Fe]\,$\simeq +0.25$ to $+0.5$ (see Table~\ref{tab:abundances})
consistent with the enhancement of the 
alpha-capture elements relative to iron seen in
halo stars of similar metallicity. 
At these low DLA metallicities, the depletion of Fe
onto dust grains is expected to be minimal 
(Vladilo 2004; Akerman et al. 2005) 
so that the observed $N$(O\,{\sc i})/$N$(Fe\,{\sc ii})
ratio gives a good indication of inherent departures from the solar
relative abundances of these two elements. As we show in 
section~\ref{sec:ion_correct}, differential ionisation
corrections between O\,{\sc i} and Fe\,{\sc ii}  are 
also unlikely to account for this effect.

\begin{table*}
\centering
\begin{minipage}[c]{0.82\textwidth}
    \caption{\textsc{Element Abundances}}
    \begin{tabular}{@{}llccccccc}
    \hline
    \hline
  \multicolumn{1}{c}{QSO}
& \multicolumn{1}{c}{$z_{\rm abs}$} 
& \multicolumn{1}{c}{$\log N$(H~{\sc i})}
& \multicolumn{1}{c}{[O/H]$^{\rm a}$}
& \multicolumn{1}{c}{$\log$\,(C/O)$^{\rm b}$}
& \multicolumn{1}{c}{$\log$\,(N/O)$^{\rm c}$}
& \multicolumn{1}{c}{[Al/H]$^{\rm a}$}
& \multicolumn{1}{c}{[Si/H]$^{\rm a}$}
& \multicolumn{1}{c}{[Fe/H]$^{\rm a}$}\\
    \hline
Q0913$+$072      & 2.61843  & 20.34 & $-2.37$ & $-0.65$ & $-2.34$      & $-2.99$ & $-2.55$ & $-2.80$  \\
SDSS\,J1016+4040 & 2.81633  & 19.90 & $-2.43$ & $-0.47$ & $\leq -1.37$ & \ldots  & $-2.51$ & \ldots   \\ 
SDSS\,J1558+4053 & 2.55332  & 20.30 & $-2.42$ & $-0.32$ & $-1.88$      & $-2.82$ & $-2.49$ & $-2.68$  \\
Q2206$-$199      & 2.07624  & 20.43 & $-2.04$ & $-0.64$ & $-2.26$      & $-2.69$ & $-2.29$ & $-2.55$  \\
                 &          &       &         &         &              &         &         &          \\
Other Objects:   &          &       &         &         &              &         &         &          \\
J0137$-$4224$^{\rm d}$  & 3.665    & 19.11 & $-2.39$ & $-0.24$ &  \ldots & \ldots  & $-2.22$ & \ldots   \\
J2155$+$1358$^{\rm e}$  & 4.21244  & 19.61 & $-1.77$ & $-0.55$ &  \ldots & $-2.12$ & $-1.87$ & $-2.13$  \\
    \hline
    \end{tabular}
    \smallskip

$^{\rm a}${[X/H]\,$= \log {\rm (X/H)}_{\rm DLA} -  \log {\rm (X/H)}_{\odot}$. Solar abundances from Asplund et al. (2005)}.\\
$^{\rm b}${These values are \textit{not} referred to the solar abundance. 
For reference, $\log {\rm (C/O)}_{\odot} = -0.27$ (Asplund et al. 2005). }\\
$^{\rm c}${These values are \textit{not} referred to the solar abundance. For reference, 
$\log {\rm (N/O)}_{\odot} = -0.88$ (Asplund et al. 2005).}\\
$^{\rm d}${P\'{e}roux et al. (2007)}.\\
$^{\rm e}${Dessauges-Zavadsky et al. (2003)}.\\
    \label{tab:abundances}
\end{minipage}
\end{table*}

\subsection{Ionisation Corrections}\label{sec:ion_correct}
As explained in section~\ref{sec:abund_meas},
the relative element abundances  listed in Table~\ref{tab:abundances}
were derived by dividing the column densities $N$(X\,{\sc n})
by $N$(H\,{\sc i}), where $N$(X\,{\sc n}) is the dominant ionisation
stage of element X in H\,{\sc i} regions.
Strictly speaking, this commonly adopted assumption only
holds true for O out of the elements considered here.
It has been known for a long time that charge-exchange reactions
keep $N$(O\,{\sc i})/$N$(H\,{\sc i}) very close to O/H for
a wide range of physical conditions (e.g. Field \& Steigman 1971).
For elements which  in H\,{\sc i} regions
are mostly singly ionised---that is, C, Al, Si and Fe
in our case---the possibility exists that some H\,{\sc ii}
gas---at the same velocity as the H\,{\sc i} and thus indistinguishable
kinematically---may contribute
to column densities of their first ions; if true, this would lead to 
an overestimate of the corresponding element 
abundances.\footnote{It is worthwhile pointing out here
that the \emph{neutral} atoms of these species, which we have
neglected, make negligible
contributions to the abundance measurements, because
even in H\,{\sc i} regions $N$(C\,{\sc i})\,$\ll N$(C\,{\sc ii}),
and similar inequalities apply to Al, Si and Fe.}
Conversely, N may be overionised relative to H, so that
even in H\,{\sc i} regions $N$(N\,{\sc i})/$N$(H\,{\sc i})\,$ < $\,(N/O) 
leading to an \emph{under}estimate of the N abundance.

Of course, one of the major motivations 
for using DLAs in abundance studies,
is the fact that when the column density of
neutral gas exceeds $N$(H\,{\sc i})\,$= 1 \times 10^{20}$\,cm$^{-2}$
such ionisation corrections
are expected to be small and comparable to other sources
of error, as concluded by a number of detailed analyses
(e.g. Viegas 1995; Vladilo et al. 2001). 
Given the importance of this point for the measurement
of the C/O ratio, however, we verified that this is indeed the 
case for the two DLAs in our small sample where column densities
of ions from H\,{\sc ii} regions could be measured in our spectra.

Specifically, we run 
a series of photoionization models with the software package {\sc cloudy}
(Ferland et al. 1998; Ferland 2000)\footnote{http://www.nublado.org/},
approximating the DLAs as slabs of constant density gas
irradiated by the metagalactic ionising background (Haardt \& Madau 2001)
and the cosmic microwave background. 
We assumed a metallicity of 1/100 solar and no differential depletions
due to dust grains (Akerman et al. 2005), but the results of the 
computations are not sensitive to these assumptions.
We varied the gas volume density from $\log [n{\rm (H)}/{\rm cm}^{-3}] = -3$ to $+3$
and stopped the calculation when the measured
column density of neutral gas in each DLA was reached.
From the {\sc cloudy} output we then calculated for each element observed
the ionization correction
\begin{equation}
{\rm IC(X)} = \log \left[ \frac{N{\rm (X)}}{N({\rm H})}\right]_{\rm
intrinsic} - \log \left[ \frac{N{\rm (X\,{\scriptstyle N})}}{N({\rm
H\,{\scriptstyle I}})}\right]_{\rm computed}
\label{eq:IC(X)}
\end{equation}
which is negative when the abundance of an element has been
overestimated by neglecting the ionized gas.

\begin{figure}
  \centering
  {\hspace*{-0.25cm}\includegraphics[angle=0,width=70mm]{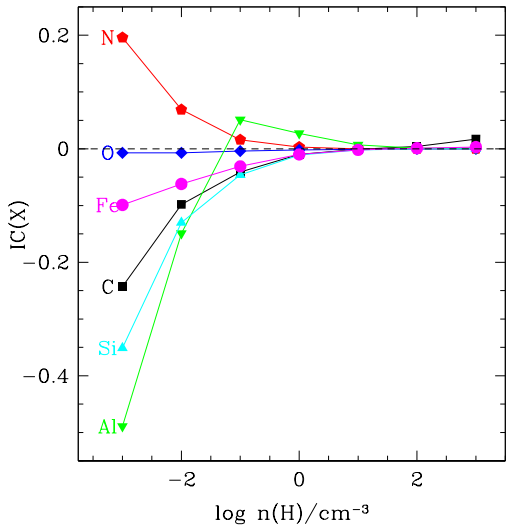}}
  \vspace{-1.5cm}
  \caption{
   Ionisation corrections (defined as in eq.~\ref{eq:IC(X)}) vs.
   gas density for observed ions which are dominant
   in H\,{\sc i} regions. The values shown here are for 
   the $z_{\rm abs} = 2.07624$ DLA in Q2206$-$199, but similar corrections 
   apply to the other DLAs in Table~\ref{tab:abundances}. 
   }
   \label{fig:cloudy_plot1}
\end{figure}

Figure~\ref{fig:cloudy_plot1} illustrates the results of this
exercise for the DLA in Q2206$-$199 (the conclusions are
similar for the other DLAs considered here).  As expected,
O\,{\sc i} tracks H\,{\sc i} very closely at all densities.
At very low densities N is overionised relative to H
(that is some N is present as N\,{\sc ii} even in H\,{\sc i}
regions), while the column densities of the 
other elements considered have some
contribution from ionised gas so that, for example,
$N$(C\,{\sc ii})/$N$(H\,{\sc i})\,$ > $\,C/H. 
However, for the elements which are the focus
of the present work, C, N (and O), the ionisation
corrections amount to less than $0.1$\,dex 
provided the gas density is greater than $n$(H)\,$ \simeq 0.01$\,cm$^{-3}$.
When $n$(H)\,$\geq 0.1$\,cm$^{-3}$, the ionisation corrections
for all the elements in Table~\ref{tab:abundances}
are $\simlt 0.05$\,dex.

\begin{figure}
  \centering
  {\hspace*{-0.25cm}\includegraphics[angle=0,width=70mm]{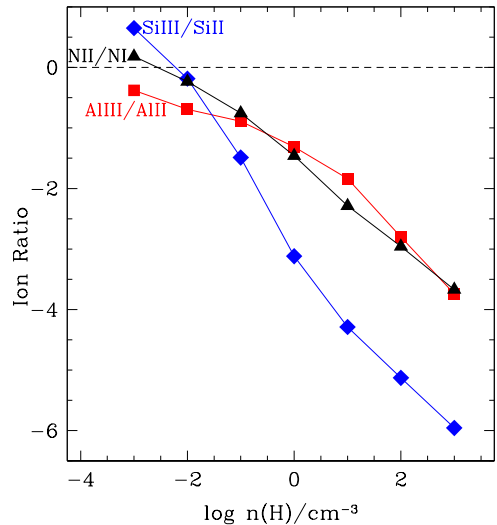}}
  \vspace{-1.5cm}
  \caption{Ratios of successive ion stages vs. gas density for the DLA in line to
   Q2206$-$199. The values shown here were calculated with a grid of
   {\sc cloudy} models (see text for details). The measured values 
   of the ion ratios (given in Table~\ref{tab:Q2206_coldens}) indicate
   $\log n{\rm (H)/cm}^{-3} \geq -2$.
   }
   \label{fig:cloudy_plot2}
\end{figure}

We can obtain an indication of the density which applies
to the gas in the DLAs from consideration of the ratios
of different ions stages, since they also depend on the 
density, once the radiation field has been fixed (see Figure~\ref{fig:cloudy_plot2}). 
The DLA which provides the most extensive information
in this respect is the one in line to Q2206$-$199. Comparing 
the measured $N$(N\,{\sc ii})/$N$(N\,{\sc i}), 
$N$(Al\,{\sc iii})/$N$(Al\,{\sc ii}), 
and $N$(Si\,{\sc iii})/$N$(Si\,{\sc ii})
ratios in Table~\ref{tab:Q2206_coldens} 
(all these ratios are $< 1$) with
the values computed by {\sc cloudy} for different
densities, we find that in this DLA
$\log n$(H)/cm$^{-3} \geq -2$ so that the ionisation
corrections for carbon and nitrogen are 
$|{\rm IC(C), IC(N)} | < 0.1$. 
This is also the case for the DLA in Q0913+072,
although in this case we rely solely on the 
$N$(Al\,{\sc iii})/$N$(Al\,{\sc ii}) ratio.
P\'{e}roux et al. (2007) and 
Dessauges-Zavadsky et al. (2003)
reached similar conclusions for the two sub-DLAs
included in Table~\ref{tab:abundances}, based 
on similar photoionisation analyses. 
For the two DLAs in the SDSS QSOs
we have no comparable information 
and therefore cannot come to definite conclusions
concerning ionisation corrections.
However, even in these two cases,
it is unlikely that to $|{\rm IC(C), IC(N)} | >  0.2$.
Larger ionisation corrections would imply
volume densities  $n$(H)\,$< 0.001$\,cm$^{-3}$
and therefore linear sizes for the absorbing structures
in excess of $\sim 100$\,kpc 
(for $N$(H\,{\sc i})\,$= 2 \times 10^{20}$\,cm$^{-2}$);
it seems unlikely that gas in such large volumes would
maintain internal velocity dispersions of only a few km~s$^{-1}$
as required by the narrow line widths.

In summary, our best estimates of the ionisation structure
of the sample of six DLAs and sub-DLAs considered here
leads us to conclude that the measured ratios
$N$(C\,{\sc ii})/$N$(O\,{\sc i}) and 
$N$(N\,{\sc i})/$N$(O\,{\sc i}) reflect
the true values  of C/O and N/O in the DLAs
to within 0.2 dex  and probably to within
0.1 dex  in most cases.

\section{The C/O ratio at low metallicities}\label{sec:C/O}

\begin{figure}
  \vspace*{-1cm}
  \centering
  {\hspace*{-0.775cm}\includegraphics[angle=0,width=100mm]{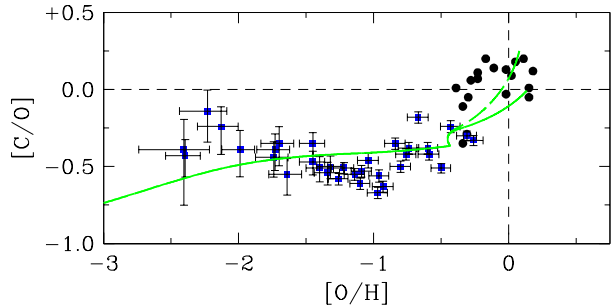}}
 \vspace{-8.25cm}
 \caption{
   [C/O] vs. [O/H] for Galactic stars from the study by Akerman et al. (2004).
   Black dots: disk stars; blue squares: halo stars. 
   The green line shows the evolution of the [C/O] ratio
   with metallicity for a `standard'  Galactic chemical evolution
   model using conventional stellar yields (see Akerman et al. 2004
   for further details). A general feature of this class of models
   is the drop in [C/O] at the lowest metallicities, as the production of
   C in massive stars lags behind that of O. The stellar
   measurements by Akerman et al. (2004) provided the first
   indication of a different behaviour, suggesting that the [C/O] ratio may
   in fact be closer to solar at metallicities [O/H]\,$\simlt -2.5$.     
   The discontinuity at [O/H]\,$= -0.5$ in the line describing the model 
   evolution corresponds to the transition from the halo to the disk.
   }
   \label{fig:co_stars}
\end{figure}

One of the aims of the present work is to
investigate the behaviour of the C/O ratio at 
low metallicities, motivated by the 
results of Akerman et al. (2004) 
and Spite et al. (2005) which we
now summarise.
Akerman et al. (2004) measured the abundances of 
C and O in 34 halo stars of metallicities from
[Fe/H]\,$ = -0.7$ to $-3.2$ and combined 
such data with analogous measurements in
disk stars from the literature. The pattern
that emerged from their study is illustrated
in Figure~\ref{fig:co_stars}.
Briefly, in halo stars [C/O]\,$\approx  -0.5$
(corresponding to an observed $\log {\rm (C/O)} \approx -0.77$
since in the Sun $\log {\rm (C/O)}_{\odot} = -0.27$; Asplund et al.
2005).\footnote{Note that in the earlier compilation 
by Grevesse \& Sauval (1998) 
the solar ratio of C and O had a similar value
$\log {\rm (C/O)}_{\odot} = -0.31$; thus the C/O ratio appears
to be robust the 3D and non-LTE
effects which have introduced significant changes to the 
individual abundances of C and O.} 
At the higher metallicities of most disk stars,
C/O rises to solar proportions; in the Galactic
chemical evolution models considered by Akerman 
et al. (2004) this rise is interpreted as the additional contribution
to carbon enrichment of the interstellar medium from the
winds of Wolf-Rayet stars ---whose mass loss rates are known to 
increase with metallicity---and, to a lesser extent, from the delayed
evolution of low and 
intermediate mass stars.

More puzzling is the apparent upturn in [C/O] 
with \emph{decreasing} metallicity when [O/H]\,$\simlt -1$,
since conventional stellar yields would predict
the opposite behaviour: the expected time lag in the production of
C relative to O in the first episodes of star formation
should cause the [C/O] ratio to decrease dramatically
as we move to the lowest metallicities in Figure~\ref{fig:co_stars}
(interested readers are
referred to the extensive
discussion of the relevant chemical evolution models 
by Akerman et al. 2004).
Akerman et al.  speculated that the trend towards
higher, rather than lower, values of [C/O] 
when [O/H]\,$< -1$ may be due to remaining traces of
high carbon production by the first stars to form
in the halo of the Milky Way proto-galaxy. Indeed,
some calculations of nucleosynthesis by Population~III
stars (e.g. Chieffi \& Limongi 2002)
do entertain high carbon yields (relative to oxygen)
that can reproduce the observations in 
Figure~\ref{fig:co_stars}.
More recently, Chiappini et al. (2006) have proposed
that the observed behaviour can also be explained if
stars of lower metallicity rotate faster: the higher
rotation speeds greatly increase the yields of
C (and N) by the massive stars which are the main
source of O.

Some uncertainty in the interpretation of the results in
Figure~\ref{fig:co_stars} is introduced by the fact that
Akerman et al. (2004) based their abundance measurements
on one-dimensional, local thermodynamical equilibrium (LTE),
stellar atmosphere models. The authors
pointed out themselves the possibility that 
the neglect of non-LTE corrections
to the high excitation C\,{\sc i} and O\,{\sc i} lines
used for abundance determinations
may mimic (or at least 
enhance) the rise in [C/O] with decreasing [O/H].
An initial investigation of such effects by
Fabbian et al. (2006) seemed to dilute
the significance of the results by Akerman et al. (2004) 
by showing that non-LTE corrections to the 
C\,{\sc i} lines can be as high as $-0.4$\,dex in stars
of [Fe/H]\,$\simeq -3$. A more recent study by the 
same group, however, has uncovered even larger
corrections to the abundance of O at very low metallicities
(Fabbian et al. 2008a), so that the 
case for a rise in [C/O] when [O/H]\,$\simlt -1$
is actually reinforced when non-LTE effects on the formation
of the stellar C\,{\sc i} and O\,{\sc i} lines are taken into account
(Fabbian et al. 2008b).

This discussion highlights the important role that 
low metallicity DLAs 
can play in clarifying the early nucleosynthetic sources
of carbon. The physics of absorption line formation
in diffuse interstellar clouds is considerably simpler
than that pertaining to stellar atmospheres; 
we do not have to take into account
the geometry of the absorbers nor their
exact thermodynamical conditions.
The only two limiting factors to the accuracy
with which we can measure the C/O ratio
are: (i) line saturation
and (ii) ionisation corrections.
In the preceding sections we have shown that
the former, if present at all in the data considered
here, would lead us to underestimate
the value of C/O,  while the latter would 
work in the opposite direction but the effect
is unlikely to be greater than 0.1\,dex
for the DLAs in the present study.

\begin{figure}
  \vspace*{-1cm}
  \centering
  {\hspace*{-0.775cm}\includegraphics[angle=0,width=100mm]{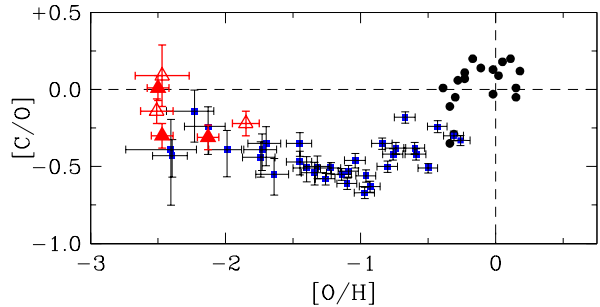}}
 \vspace{-8.25cm}
  \caption{
   The values of [C/O] deduced here for DLAs (red filled triangles)
   and sub-DLAs (open triangles) broadly agree with the stellar abundances
   measured at similar metallicities, supporting the suggestion by
   Akerman et al. (2004) that carbon and oxygen may be produced
   in near-solar proportions in the earliest stages of galactic chemical evolution.  
   Other symbols have the same meaning as in Figure~\ref{fig:co_stars}.
   }
   \label{fig:co_DLAs}
\end{figure}

When the values of [C/O] measured in the six
DLAs/sub-DLAs in the present sample
are compared with those of Galactic stars
as in Figure~\ref{fig:co_DLAs},  it appears
that both sets of data paint a consistent
picture.\footnote{The values of [C/O] plotted
in Figure~\ref{fig:co_DLAs} were deduced
from the values of $\log {\rm (C/O)}$
listed in Table~\ref{tab:abundances} after subtracting
the solar value  $\log {\rm (C/O)}_{\odot} = -0.27$
from the compilation by Asplund et al. (2005).
The errors shown were obtained by combining in
quadrature the $1 \sigma$ errors in the 
column densities $N$(C\,{\sc ii}) and 
$N$(O\,{\sc i}) returned by VPFIT and listed 
in Tables~\ref{tab:Q0913_coldens}, \ref{tab:J1016_coldens}, 
\ref{tab:J1558_coldens} and \ref{tab:Q2206_coldens}, 
or given by the original authors for the two
DLAs from the literature. In all cases, a minimum 
column density uncertainty
of 0.05\,dex was assumed.}
The DLA measurements corroborate
the conclusion by Akerman et al. (2004) in 
showing relatively high values of C/O 
in the low metallicity regime and complement
the still very limited statistics of this
ratio in stars of metallicity [O/H]\,$< -2$.

Although the high redshift DLAs and the halo stars
appear to constitute a continuous sequence in 
Figure~\ref{fig:co_DLAs}, we warn against such
a simplistic interpretation. The chemical enrichment 
histories of the Milky Way halo and of the galaxies
giving rise to DLAs need not be the same. Therefore
it may well be that, as chemical evolution
progresses, different DLAs follow different
paths in the  [C/O] vs. [O/H] plane, some
similar to that of the Milky Way stellar populations
while others may be different. Such a variety of
behaviours has already
been seen (in other elements) in Local Group
galaxies (e.g. Pritzl et al. 2005).
What our data show is that at metallicities
below 1/100 of solar DLAs and Galactic stars 
concur in showing elevated ratios of C/O.
Akerman et al. (2004) speculated that an 
approximately solar value of C/O
might be recovered when [O/H]\,$\simlt -3$;
the new data presented here are certainly
consistent with this extrapolation of the Galactic
measurements.

\section{The N/O ratio at low metallicities}

In contrast with the paucity of carbon measurements,
several studies have targeted the abundance of N in 
DLAs since the first determination of the N/O ratio
in a metal-poor DLA by Pettini, Lipman, \& Hunstead (1995). 
These authors pointed out the potential of 
such data for clarifying the nucleosynthetic origin of N
and unravelling the chemical enrichment history of 
the galaxies giving rise to damped systems.
Henry \& Prochaska (2007) and Petitjean, Ledoux, \& Srianand (2008)
provide recent
discussions of these issues and comprehensive 
reviews of earlier literature.

At present, there are 35 DLAs with reliable measurements (or useful
upper limits) of N/O, collected in Table~\ref{tab:no}.\footnote{In most of
these cases S is used as a proxy for O because, unless far-UV O\,{\sc i}
lines are accessible (and these are often blended in the \lya\ forest),
the strong O\,{\sc i}\,$\lambda 1302$ line is saturated in most
DLAs. The use of S as a proxy for O is justified by the fact that
S behaves like a typical alpha-capture element in Galactic stars
(e.g. Nissen et al. 2007).}
Despite this body of data, it is still instructive to consider
the three measurements of N/O reported
here, as they refer to some of the most metal-poor DLAs known.
In order to understand their importance, it is necessary
to summarise current ideas on the production of N in stars
of different masses---we refer the interested reader to 
Henry \& Prochaska (2007), Pettini et al. (2002a), and references
therein for more detailed treatments.

\begin{table*}
\centering
\begin{minipage}[c]{0.99\textwidth}
    \caption{\textsc{N, O, S, and Fe Abundance Measurements in DLAs}}
    \begin{tabular}{@{}lrrrrrrrrr}
    \hline
    \hline
   \multicolumn{1}{c}{QSO}
& \multicolumn{1}{c}{$z_{\rm abs}$} 
& \multicolumn{1}{c}{$\log N$\/(H\,{\sc i})}
& \multicolumn{1}{c}{$\log N$\/(N\,{\sc i})}
& \multicolumn{1}{c}{$\log N$\/(O\,{\sc i})}
& \multicolumn{1}{c}{$\log N$\/(S\,{\sc ii})}
& \multicolumn{1}{c}{$\log {\rm (O/H)}+12^a$}
& \multicolumn{1}{c}{$\log {\rm (N/O)}^a$}
& \multicolumn{1}{c}{$\log {\rm (Fe/O)}^a$}
& \multicolumn{1}{c}{Ref.$^b$}\\
    \multicolumn{1}{c}{}
& \multicolumn{1}{c}{}
& \multicolumn{1}{c}{(cm$^{-2}$)}
& \multicolumn{1}{c}{(cm$^{-2}$)}
& \multicolumn{1}{c}{(cm$^{-2}$)}
& \multicolumn{1}{c}{(cm$^{-2}$)}
& \multicolumn{1}{c}{}
& \multicolumn{1}{c}{}
& \multicolumn{1}{c}{}
& \multicolumn{1}{c}{}\\    
  \hline
Q0000$-$2620    & 3.3901   & 21.41 & 14.73           & 16.45    & \ldots   & 7.04 & $-1.72$ &   $-1.58$ &   1 \\
Q0100+1300      & 2.30903  & 21.37 & 15.03           & \ldots   & 15.09    & 7.22 & $-1.56$ &   $-1.50$ &   2 \\
Q0112$-$306     & 2.41844  & 20.50 & 13.16           & 14.95    & \ldots   & 6.45 & $-1.79$ &   $-1.62$  &   3 \\
Q0201+1120      & 3.38639  & 21.26 & 15.33           & \ldots   & 15.21    & 7.45 & $-1.38$ &   $-1.36$ &   4 \\
J0307$-$4945    & 4.46658  & 20.67 & 13.57           & 15.91    & \ldots   & 7.24 & $-2.34$ &   $-1.70$ &    5 \\
Q0347$-$3819    & 3.02485  & 20.73 & 14.64           & 16.45    & 14.74   & 7.72 & $-1.81$ &   $-2.10$ &   3 \\
Q0407$-$4410    & 2.5505   & 21.13 & 14.55           & \ldots   & 14.82    & 7.19 & $-1.77$ &   $-1.37$ &   6 \\
Q0407$-$4410    & 2.5950   & 21.09 & 15.07           & \ldots   & 15.19    & 7.60 & $-1.62$ &   $-1.54$ &   6\\
Q0528$-$2505    & 2.1410   & 20.95 & 14.58           & \ldots   & 14.83    & 7.38 & $-1.75$ &   $-1.48$ &   7 \\
HS0741+4741     & 3.01740   & 20.48 & 13.98          & \ldots   & 14.01    & 7.03 & $-1.53$ &   $-1.46$ &   8 \\
0841+1256       & 2.37452  & 20.99 & 14.60           & \ldots   & 14.69    & 7.20 & $-1.59$ &   $-1.43$ &   9 \\
0841+1256       & 2.47621  & 20.78 & 13.94           & 16.15    & 14.48    & 7.37 & $-2.21$ &   $-1.65$ &   9 \\
J0900+4215      & 3.2458   & 20.30 & 14.15           & \ldots   & 14.65    & 7.85 & $-2.00$ &   $-1.61$ &   8 \\
Q0913+072       & 2.61843  & 20.34 & 12.29           & 14.63    & \ldots   & 6.29 & $-2.34$ &   $-1.64$ &   10 \\
Q0930+2858      & 3.2353   & 20.18 & 13.82           & \ldots   & 13.67    & 6.99 & $-1.35$ &   \ldots  &   11 \\
Q1108$-$077    & 3.60767  & 20.37 & $\leq 12.84$ & 15.37    & \ldots  & 7.00 & $\leq -2.53$  & $-1.49$ & 3 \\
Q1210+17        & 1.89177  & 20.63 & 14.71           & \ldots   & 14.96    & 7.83 & $-1.75$ &   $-1.45$ &   9 \\
Q1232+0815      & 2.33772  & 20.80 & 14.63           & \ldots   & 14.83    & 7.53 & $-1.70$ &   $-1.62$ &   7 \\
Q1331+170       & 1.77637  & 21.14 & $\leq 15.23$    & \dots    & 15.08    & 7.44 & $\leq -1.35$ & $-1.95$ & 2 \\
Q1337+113       & 2.79581  & 21.00 & 13.99          & 15.74       & \ldots    & 6.74 & $-1.75^{c}$ & $-1.41$ & 3 \\
Q1409+095       & 2.45620  & 20.54 & $\leq 13.19$    & 15.15    & \ldots   & 6.61 & $\leq -1.96$ & $-1.41$ & 12 \\
J1435+5359      & 2.3427   & 21.05 & 14.67           & \ldots   & 14.78    & 7.23 & $-1.61$ &   \ldots  & 8 \\ 
J1443+2724      & 4.224    & 20.95 & 15.52           & \ldots   & 15.52    & 8.07 & $-1.50$ &   $-1.69$ & 13 \\
SDSS\,J1558$-$0031   & 2.70262  & 20.67 & 14.46      & 15.87    & 14.07    & 7.20 & $-1.41$ &   $-1.76$ & 14 \\
SDSS\,J1558+4053     & 2.55332  & 20.30 & 12.66      & 14.54    & \ldots   & 6.24 & $-1.88$ &   $-1.47$ & 10 \\
GB1759+7539     & 2.62530  & 20.76 & 15.11           & \ldots   & 15.24    & 7.98 & $-1.63$ &   $-1.66$ & 8 \\
Q2059$-$360     & 3.08293   & 20.98 & 13.95          & 16.09       & 14.41   & 7.11 & $-2.14$ &   $-1.61$ & 3 \\ 
Q2206$-$199     & 2.07624  & 20.43 & 12.79           & 15.05    & \ldots   & 6.62 & $-2.26$ &   $-1.72$ &   10 \\
Q2230+02        & 1.8644   & 20.85 & 15.02           & \ldots   & 15.29    & 7.94 & $-1.77$ &   $-1.30$ &   9, 8 \\
Q2231$-$002     & 2.06616  & 20.53 & $\leq 15.02$    & \dots    & 15.10    & 8.07 & $\leq -1.58$ & $-1.77$ & 2 \\
HE2243$-$6031   & 2.33000  & 20.67 & 14.88           & \ldots   & 15.02    & 7.85 & $-1.64$ &   $-1.60$ &   15 \\
Q2332$-$094     & 3.05723   & 20.50  & 13.73           & 15.95     & 14.37   & 7.45  & $-1.94^{c}$ & $-1.58$ & 3 \\
Q2342+342       & 2.9082   & 21.10 & 14.92           & \ldots   & 15.19    & 7.59 & $-1.77$ &   $-1.67$ &   8 \\
Q2343+1232      & 2.43125  & 20.35 & 15.22           & \ldots   & 14.86    & 8.01 & $-1.14$ &   $-1.70$ &   16 \\
Q2348$-$1444    & 2.27939  & 20.59 & 13.35           & \ldots   & 13.75    & 6.66 & $-1.90$ & $-1.41$ & 9 \\
    \hline
    \end{tabular}
    \smallskip

$^{\rm a}${When the oxygen abundance
is not available, S has been used as a proxy for O
by assuming the solar ratio $\log {\rm (O/S)}_{\odot} = +1.50$ (Asplund et al. 2005).}\\
$^{\rm b}${References---1: Molaro et al. (2000); 
2: Dessauges-Zavadsky et al. (2004);
3: Petitjean, Ledoux, \& Srianand (2008);
4: Ellison et al. (2001);
5: Dessauges-Zavadsky et al. (2001);
6: Lopez \& Ellison (2003);
7. Cent\'{u}rion et al. (2003);
8. Henry \& Prochaska (2007).
9. Dessauges-Zavadsky et al. (2006);
10: This work;
11. Lu, Sargent, \& Barlow (1998);
12. Pettini et al. (2002a).
13. Ledoux et al. (2006b).
14. O'Meara et al. (2006).
15: Lopez et al. (2002); 
16: D'Odorico, Petitjean, \& Cristiani (2002).
}\\
$^{\rm c}${(N/O) ratio only for components in common between N\,{\sc i} and O\,{\sc i}.}
\\
    \label{tab:no}
\end{minipage}
\end{table*}

\begin{figure*}
  \vspace*{-2.95cm}
  \centering
  {\hspace*{-0.75cm}\includegraphics[angle=270,width=190mm]{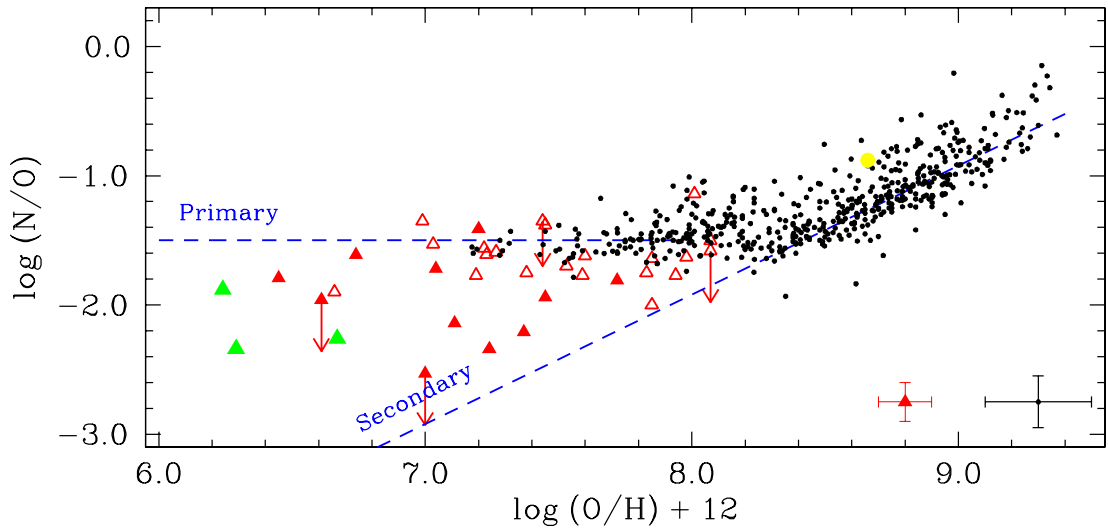}}
  \vspace{-3.25cm}
  \caption{Abundances of N and O in extragalactic H\,{\sc ii} regions
  (small dots) and high redshift damped \lya\ systems (triangles). 
  The former are from the sources assembled by Pettini et al. (2002a), 
  with the addition 
  of more recent data from Melbourne et al. (2004).
  The DLA values are from the compilation in Table~\ref{tab:no};
  filled triangles denote DLAs where the abundance of O could be measured
  directly, while open triangles are cases where S was used as a proxy for O.
  The three green filled triangles are the new determinations
  reported in this paper.
  The error bars in the bottom right-hand corner
  give an indication of the typical uncertainties;
  the large yellow dot corresponds to the 
  solar abundances of N and O (Asplund et al. 2005).
  The dashed lines are approximate representations
  of the primary and secondary levels of N production (see text).  }
   \label{fig:no_DLAs}
\end{figure*}

The key points are as follows. 
There is general
agreement that the main pathway for the production of
N in stellar interiors is a six step process in the CN
branch of the CNO cycle which takes place in the stellar H
burning layer, with the net result that $^{14}$N is synthesised
from $^{12}$C and $^{16}$O. Less well established, however, are 
the N yields of stars of different masses and metallicities.
The issue is complicated by the fact that N can be 
of either `primary' or `secondary' origin,
depending on whether the seed carbon and
oxygen are those manufactured by the star during helium burning,
or were already present when the star first condensed out of the 
interstellar medium (ISM).

This dual nature is revealed when the abundances of N and O
in extragalactic H\,{\sc ii} regions (these are the main
source of measurements of these two elements in the nearby universe)
are compared as in Figure~\ref{fig:no_DLAs}.\footnote{See Pettini et al. (2002a)
for the sources of the H\,{\sc ii} region measurements reproduced in
Figure~\ref{fig:no_DLAs}. Although much more extensive data sets
have become available recently through large scale surveys such as the 
Sloan Digital Sky Survey (e.g. Mallery et al. 2007; Liang et al. 2007---see
also Nava et al. 2006),
the locus occupied by H\,{\sc ii}
regions in the log~(N/O) vs. log~(O/H) plane has not changed significantly.}
When the oxygen abundance is greater than about half solar,
that is when $\log \rm {(O/H)} + 12 \simgt 8.3$ adopting 
$\log \rm {(O/H)}_{\odot} + 12 = 8.66$ from Asplund et al. (2005),
N/O rises steeply with increasing O/H; this is the regime where
N is predominantly secondary.\footnote{Henry, 
Edmunds, \& K\"{o}ppen (2000) pointed out that the rise
in N/O with O/H is steeper than would be normally expected
for a purely secondary element and proposed that it is augmented
by a decreasing O yield with increasing metallicity.}
At lower metallicities on the other hand 
($\log {\rm (O/H)}+12~\simlt 8.0$), N is mostly primary 
and its abundance tracks that of O; this results in the flat
plateau at $\log {\rm (N/O)} \simeq -1.5$ evident 
in Figure~\ref{fig:no_DLAs}.

Primary N is thought to be synthesised most effectively 
by intermediate mass stars on the asymptotic giant branch;
consequently, its release into the interstellar medium
presumably takes place some time after 
the massive stars which are the main source of oxygen
have exploded as Type~II supernovae. 
Henry et al. (2000) calculated the delay to be 
$\Delta t \sim 250$\,Myr for a `standard' stellar initial mass function
(IMF) and published stellar yields.

Damped \lya\ systems add information that can be of value
in assessing the validity of this overall picture. First of all,
they offer measures of  the abundances  of O and N in entirely
different environments from nearby star-forming galaxies.
Second, they sample low metallicity regimes which are rare
today. And third, by virtue of their being at high redshift,
they can in principle provide a better sampling of the time
delay in the production of primary N relative to O---the 
canonical $\Delta t \sim 250$\,Myr is a much
larger fraction of the time available for star formation at
$z = 2 - 3$ than today. In other words, the chances 
of `catching'  a galaxy in the interim period
following a burst of star formation, when the O from
Type~II supernovae has already dispersed into the
interstellar medium but lower mass stars have yet to 
release their primary N, are considerably higher in 
DLAs than in metal-poor star-forming galaxies in the
local universe. 
Such a situation should manifest itself as a displacement
below the primary plateau in the log~(N/O) vs. log~(O/H) plane
in Figure~\ref{fig:no_DLAs}.

The number of DLAs with reliable measures of the N and O
abundances has slowly grown since the first determination
by Pettini et al. (1995); 
the current total of 35 such systems (Table~\ref{tab:no})
provides a moderate size sample with which to confront
the above scenario. 
When DLAs are compared with present-day galaxies, 
as in Figure~\ref{fig:no_DLAs},
it can be seen that in most cases their values of 
N/O fall close to the plateau defined by
metal-poor H\,{\sc ii} regions.
If primary N and O are synthesised by stars of
different masses, as generally believed, the good
agreement between local and distant galaxies
attests to the universality of the IMF---a top-heavy
IMF in DLAs would result in lower values of N/O,
contrary to observations. 

Out of the 35 DLAs in our sample, 12 ($\sim 34$\%)
lie more than 0.3\,dex 
[twice the typical $1 \sigma$ error in the determination
of $\log{\rm (N/O)}$]
below the primary plateau.
For comparison, $\Delta t / [1/H(z)] \simeq 0.18$ at
the mean $\langle z \rangle  = 2.70$ of the 35 DLAs in Table~\ref{tab:no},
where $1/H(z)$ is the age of the universe at redshift $z$
(in today's consensus cosmology). 
The difference between these two fractions
is probably not significant, 
given that the DLA sample in Table~\ref{tab:no}
is somewhat biased towards chemically 
unevolved DLAs and that the age of the universe is 
in any case an upper
limit to the time available for star formation. 
Certainly, the current DLA statistics are consistent with
a time delay $\Delta t$ of a few $10^8$ years and allay earlier
concerns based on more limited samples (e.g. Pettini et al. 2002a).

Figure~\ref{fig:no_DLAs} also emphasises the importance
of extending measurements of N/O to the lowest
metallicity DLAs: as the gap between secondary and
primary N production widens, it becomes easier to recognise
additional sources of primary nitrogen, if they exist.
That is, if intermediate mass stars are the sole producers
of primary N, it should be possible to find DLAs with
values of N/O close to the secondary line in 
Figure~\ref{fig:no_DLAs}. 
On the other hand, if massive stars were able to synthesise
\emph{some} primary N, we might expect to see
a minimum value
of N/O, corresponding to the N ejected into the ISM
at the same time as O. 
This idea was first proposed by
Centuri{\'o}n et al. (2003---see also Molaro 2003) and recent
work on stellar yields including the effects of rotation
(e.g.  Hirschi et al. 2007) suggests that it may be plausible.
In this respect, the new data presented here 
(shown by the green triangles in Figure~\ref{fig:no_DLAs})
add some weight to the suggestion by 
Centuri{\'o}n et al. (2003): the lowest metallicity DLAs 
show values of N/O as low as, but no lower than,
the lowest previously measured 
(in more metal-rich DLAs).
While the statistics are still very limited, 
the available data are at least consistent
with a `floor' in the value of N/O at 
$\log {\rm (N/O)} \simeq -2.3$.
If this does indeed reflect the primary N yield
by massive stars, it would imply that 
at these metallicities massive stars 
can account for $\sim 15$\%
of the total primary N production.
Clearly, it will be of interest to test the validity
of this interpretation as more metal-poor DLAs are
studied at high spectral resolution and the sample
of N/O measurements at [O/H]\,$< -2$
increases in the years ahead.

\section{Summary and Conclusions}
In this paper we have presented new observations of four
damped absorption systems (three DLAs and one sub-DLA
according to the conventional division between the two
at $\log N$(H\,{\sc i})\,$= 20.3$) which are among the
least metal-enriched known. 
We have focussed in particular on the clues
which the abundance patterns of these DLAs 
provide on the nucleosynthesis of the C, N, O 
group at low metallicities. 
Our main conclusions are as follows.

\begin{itemize}

\item[1.] DLAs at the low end of the metallicity distribution
can be preselected from medium resolution QSO spectra
by targeting damped \lya\ lines which apparently 
have no associated metal absorption lines.
As expected, such systems turn out to have \emph{narrow}
metal lines when observed at higher spectral resolution,
consisting of only a few components spread out over a few
tens of km~s$^{-1}$.

\item[2.] The most metal-poor DLAs known have
oxygen abundances of $\sim 1/250$ solar 
([O/H]\,$\simeq -2.4$), and `metallicities'
which are lower by factors of
$\sim 2 - 3$ when measured in carbon and iron.
This level of chemical enrichment is similar
to that typical of intergalactic \lya\ forest
clouds and suggest that these DLAs have experienced
very little, if any, local star formation activity.
No DLAs have yet been discovered with 
[Fe/H]\,$< -3$, comparable to the iron abundance
of the most metal-deficient stars in the Galactic halo,
but such stars are very rare and larger samples
of DLAs may be required to intersect such pristine
clouds of neutral gas at high redshift.

\item[3.] Together with two sub-DLAs from earlier
published studies, we have assembled a small
sample of six absorption systems where the 
C/O ratio can be measured with confidence.
We find good agreement between the relatively
high values of C/O in these metal-poor DLAs
and those measured by Akerman et al. (2004)
in halo stars of similar oxygen abundance. 
The DLA 
determinations are less ambiguous than their
stellar counterparts. 
The agreement we have found 
suggests that corrections
to the stellar abundances of C and O
resulting from more sophisticated treatments
of the stellar atmospheres (including 3-D and non-LTE
effects) are likely to be of similar magnitude
for the lines analysed by Akerman et al. (2004),
so that the net corrections to the values
of C/O they derived may not be large.
The apparent rise of the C/O ratio as we move
from stars with [O/H]\,$\simeq -1$ to lower
metallicities points to an additional source of
carbon from the massive stars responsible for
early episodes of nucleosynthesis, in the Milky Way
halo and in high redshift DLAs. 

\item[4.] The data presented here extend the
growing sample of N/O measurements
in DLAs to lower oxygen abundances, where
the distinction between secondary and primary
channels widens. Our results strengthen earlier
suggestions of a `floor' in the value of N/O
at $\log {\rm (N/O)} \simeq -2.3$ 
(or $\sim 1/25$ of the solar N/O ratio),
although larger samples of metal-poor DLAs 
are necessary to establish its reality. 
If this floor reflects the small primary
production of  N by massive stars,
it amounts to $\sim 15$\% of the total
primary N yield integrated along the IMF.
If most of the primary N is from intermediate
mass stars, the generally good agreement in
N/O between most DLAs and nearby low
metallicity galaxies is an indication that the 
IMF does not vary significantly between 
these different environments and cosmic 
epochs. 
More generally, our results add weight
to the universality of 
the stellar yields of C, N, and O (Henry 2004),
given the consistency in the ratios
C/O and N/O between high redshift DLAs,
Galactic stars, and nearby H\,{\sc ii} regions.

\end{itemize}

What is still unclear is where and when these
trace metals, which account for less than 
$\sim 5 \times 10^{-5}$ of the baryons by
mass, were synthesised. 
The element ratios in the most metal-poor
DLAs offer clues which may also be relevant
to the origin of the metals 
polluting the intergalactic medium (IGM),
where abundances are less well determined
(as they rely on photoionisation models)
but appear to be similar to those deduced here.

In particular, it is now well established that
star-forming galaxies at $z = 2 - 3$ drive large-scale
gas outflows into their surroundings
(Adelbeger et al. 2005; Steidel et al. in preparation);
the metallicity of the outflowing material is likely
to be within a factor of a few of solar (Pettini et al.
2002b; Erb et al. 2006a).
A novel suggestion recently put forward by
Schaye, Carswell, \& Kim (2007) is that
the low metallicities commonly deduced 
for the IGM are misleading, resulting from
the poor mixing between small parcels
of ejected gas of near-solar composition
and true intergalactic gas of primordial origin
on much larger scales.
The relatively high C/O values we have 
uncovered in the DLAs studied here \emph{could}
be viewed in this context, as they are not
dissimilar to those also measured in disk stars
with [O/H]\,$\simgt -0.5$ (see Figure~\ref{fig:co_DLAs}). 
In this case, the good agreement with metal-poor halo
stars with [O/H]\,$\simlt -2$ would be coincidental.
A more extensive set of C/O measurements in both
stars and DLAs should clarify the situation; for example,
a definite \emph{trend} of increasing C/O with decreasing
[O/H] would not be expected on the basis of the model
proposed by Schaye et al. (2007).  

The few DLAs with the lowest values of N/O are 
more discriminatory in this respect. The median ages
of the galaxies which support the outflows at $z = 2 - 3$ 
are between $\sim 300$ and $\sim 600$\,Myr (Shapley et al. 2001;
Papovich et al. 2001; Erb et al. 2006b), comparable to, or 
somewhat larger than, the time delay $\Delta t$ for the release of
primary N from intermediate mass stars. But even in the 
younger galaxies, the secondary N produced when the
metallicity is high ([O/H]\,$\simgt -0.5$, Erb et al. 2006a) would result
in N/O ratios significantly higher than observed in the
most metal-poor DLAs. 
For example, in the young
Lyman break galaxy MS1512-cB58, where [O/H]\,$\simeq -0.4$
(Teplitz et al. 2000),
Pettini et al. (2002b)
deduced $\log {\rm (N/O)} = -1.8$, a factor of $\sim 3$ higher
than the `floor' at $\log {\rm (N/O)} \simeq -2.3$
exhibited by DLAs.  Taken together, the C/O and N/O
measurements reported here favour an origin of these
elements in stars of genuinely low metallicity, rather than
in diluted ejecta from galaxies which have already attained
high levels of chemical enrichment.

In this case, our data hint to the possibility
that massive stars may contribute significantly 
to the nucleosynthesis
of both C and N, as well as O, at the lowest
metallicities. Stellar rotation may be the common
factor here, through the mixing it induces in the
stars' interiors, but it remains to be established if
physically motivated metallicity-dependent 
C, N, and O yields can reproduce the 
relative abundances of these elements measured in 
DLAs, when incorporated
into appropriate chemical evolution models.
From an observational point of view, it should be
possible with current instrumentation and 
some dedicated effort to increase the numbers
of absorption systems with reliable measures 
of C/O and N/O and place the conclusions
reached here on stronger statistical footing.
We look forward to such developments in the
years ahead.

\section*{Acknowledgments}
We are grateful to the Keck and ESO 
time assignment committees for
their support of this programme and to the 
staff astronomers at the Keck and VLT 
observatories for their help in conducting the 
observations.
Tom Barlow, Bob Carswell, Thiago Gon\c{c}alves,
Michael Murphy, and Sam Rix kindly 
helped with different aspects of the data reduction
and analysis.
Martin Asplund, Damian Fabbian, Dick Henry, Poul Nissen
and an anonymous referee 
made valuable suggestions which improved the paper.
We thank the Hawaiian
people for the opportunity to observe from Mauna Kea;
without their hospitality, this work would not have been possible.
CCS's research is partly supported by grant
AST-0606912 from the US National Science Foundation.
Some of this work was carried out during a visit
by Fred Chaffee to the Institute of Astronomy, Cambridge,
supported by the Institute's visitor grant.

\end{document}